\documentclass[journal]{IAENGtran}
\usepackage{amsfonts, amssymb}

\ifCLASSINFOpdf
   \usepackage[pdftex]{graphicx}
   \DeclareGraphicsExtensions{.pdf,.jpeg,.png}
\else
   \usepackage[dvips]{graphicx}
   \DeclareGraphicsExtensions{.eps}
\fi
\usepackage[cmex10]{amsmath}

\begin{document}
%
\title{Estimating Coefficients of Frobenius Series by\\
Legendre Transform and WKB Approximation
}
%
%
%

\author{Amna Noreen and K{\aa}re~Olaussen
 \thanks{Manuscript received Mars 18, 2012.}
 \thanks{A.~Noreen is with the Department
 of Physics, NTNU, N-7048 Trondheim, Norway.
 e-mail: Amna.Noreen@ntnu.no.}
 \thanks{K.~Olaussen is with the Department
 of Physics, NTNU, N-7048 Trondheim, Norway.
 e-mail: Kare.Olaussen@ntnu.no.}}

\maketitle

\pagestyle{empty}
\thispagestyle{empty}

\begin{abstract}
The Frobenius method can be used to represent solutions
of ordinary differential equations by (generalized) power
series. It is useful to have prior knowledge of the coefficients
of this series. In this contribution we demonstrate that the magnitude
of the coefficients can be predicted to surprisingly high accuracy
by a Legendre transformation of WKB approximated solutions to
the differential equations.
\end{abstract}

\begin{IAENGkeywords}
Second order ODEs, Regular singular points, Frobenius method,
Legendre transformation, WKB approximation.
\end{IAENGkeywords}

%
\IAENGpeerreviewmaketitle

\section{Introduction}
%
%
%
%
\IAENGPARstart{R}{ecently} we have developed and used code for solving
ordinary Frobenius type differential equations to very
high precision \cite{CPC2011:AsifAmnaKareIngjald, ICCP2011:AmnaKare, CPC2012:AmnaKare},
like finding the lowest eigenvalue of
\begin{equation}
   -\psi''(x) + x^4 \psi(x) = \varepsilon \psi(x) 
\end{equation}
to one million decimals. The general class of equations
treated in \cite{CPC2012:AmnaKare} is of the type
{\footnotesize
\begin{equation}
     -\!\left(\frac{d^2}{dz^2}  + 
     \frac{1\!-\!\nu_+ \!-\! \nu_-}{z}\frac{d}{dz} + 
     \frac{\nu_+ \nu_-}{z^2} \right)\psi(z) + 
   \frac{1}{z}\sum_{n=0}^{N} \text{v}_n\, z^n\,\psi(z) = 0.
   \label{ODE}
\end{equation}
}
Following the Frobenius method our solution is represented by
a convergent series
\begin{equation}
  \psi(z) = \sum_{m=0}^\infty a_m\,z^{m+\nu},
  \label{PowerSeries}
\end{equation}
where the coefficients $a_m$ is generated recursively in parallel
with a brute force summation of the series. The individual terms 
in (\ref{PowerSeries}) may grow very big, leading to huge cancellations
and large roundoff errors. It is therefore useful to have some prior
knowledge of the magnitude of the $a_m$'s before a 
high-precision evaluation --- to set the computational precision
required for a desired accuracy of the final result, and to
estimate the time required to complete the computation.

We have found that $\vert a_m\vert$ can be
estimated surprisingly accurate from a WKB approximation
of the solution, followed by a Legendre transform. For the
general class of equations~(\ref{ODE}) the WKB integrals
and the Legendre transform must be done by (ordinary precision)
numerical methods.

In the remainder of this paper we first derive a Legendre transform
relation between the magnitudes $\vert a_m \vert$ and
$\vert \psi(z) \vert$, slightly generalized to take into account
a logarithmic correction, and next use the WKB approximation to
estimate $\psi(z)$.

\section{Legendre transform method of solution}

Our method of solution is based on the hypothesis that
the sum~(\ref{PowerSeries}) for large $\vert z \vert$ 
receives its main contribution from a
relatively small range of $m$-values.
Introduce quantities $u$ and $s(m)$ so that
\begin{equation*}
  x = \text{e}^u,\quad \vert a_m \vert = \text{e}^{s(m)}.
\end{equation*}
Our assumption is that
\begin{equation}
   \text{e}^{S(u)} \equiv \max_{\varphi} \psi(\text{e}^{u+i\varphi}) \approx 
   \sum_{m} \text{e}^{s(m)+(\nu+m)u},
   \label{MaximumAbsoluteValue}
\end{equation}
with the main contribution to the sum coming from a small range
of $m$-values around a maximum value $\bar{m}$. The latter is
defined so that $s'(\bar{m})+u=0$, $s''(\bar{m})<0$.
Now write $m = \bar{m} + \Delta m$, and approximate the
sum (\ref{MaximumAbsoluteValue}) over $\Delta m$ by a gaussian integral.
This gives
\begin{equation*}
   \text{e}^{S(u)} \approx   \sqrt{{-2\pi}/{s''(\bar{m})}}\,\text{e}^{s(\bar{m})+(\nu+\bar{m})u}.
\end{equation*}
In summary, we have found the relations
\begin{align}
    u    &= -s'(m)\label{VariableChange},\\
    S(u) &= s(m) - (\nu + m) s'(m) +\frac{1}{2}\log\left(\frac{2\pi}{-s''(m)} \right)\nonumber\\
         &\equiv S_0(u) + \frac{1}{2}\log\left(\frac{2\pi}{-s''(m)} \right).\label{FunctionChange}
\end{align}
This is essentially a Legendre transformation between $s(m)$ and $S(u)$. 
Consider a small change $u\to u+\delta u $. To maintain the maximum condition
we must also make a small change $m \to m+\delta m$, with
\(
    \delta m = -{\delta u}/{s''(m)}
\). I.e.~$s''(m)=-u'(m)$.
This is consistent with the result of taking the $m$-derivative of
equation~(\ref{VariableChange}).
One further finds that $S_0(u)$ becomes
\begin{align*}
    &S_0(u+\delta u) = S_0(u) + S'_0(u)\,\delta u + \frac{1}{2}S''_0(u)\,\delta u^2 +\cdots\\
    &=s(m) + (\nu + m) u + (m+\nu)\,\delta u - \frac{1}{2s''(m)}\, \delta u^2 + \cdots,
\end{align*}
giving the relations
\begin{align}
    (m+\nu) &= S'_0(u), \label{mExpression}\\
    s(m)    &=  S_0(u) - u S'_0(u),\label{sExpression}\\
    s''(m)  & = -S''_0(u)^{-1}. \label{SecondDerivative}
\end{align}
Equation (\ref{SecondDerivative}) just says that $\left(dm/du\right) = \left(du/dm\right)^{-1}$.
We are only able to compute $S(u)$ directly, not $S_0(u)$. However, they only differ by
a logarithmic term, hence we will approximate $\log(-s''(m)) = -\log S''_0(u) \approx -\log S''(u)$.
This gives
\begin{equation}
    S_0(u) \approx S(u) - \frac{1}{2}\log\left(2\pi\, S''(u)\right),\label{logCorrectedS0}
\end{equation}
which can be used in equations (\ref{mExpression}--\ref{SecondDerivative})
when we have computed $S(u)$.

\section{WKB approximation} 

It remains to find $S(u)$. Here we will use the leading order WKB 
approximation to find a sufficiently accurate estimate.
When $z=0$ is an ordinary point, i.e. when $\nu_-=0$, $\nu_+=1$, 
the leading order WKB solution to (\ref{ODE}) is
\begin{equation}
   \psi(z) \approx \sqrt{Q_0/Q(z)}
   \exp\left({\frac{1}{s}}\int_0^{z} Q(t) \text{d}t \right),
   \label{WKBApproximation}
\end{equation}
where $Q^2(z) = \sum_{n=1}^N \text{v}_n z^{n-1}$, and $Q_0=Q(0)$.
This represents a superposition of the solutions $\psi_\pm(z)$. 
The difference between the $\nu_+$ and $\nu_-$ solutions is at worst
comparable to accuracy of our approximation; hence we will not
distinguish between them.

When $z=0$ is a regular singular point we use the Langer corrected
WKB approximation to obtain leading order solutions in the form
\begin{align}
  \psi_{\pm}(z) &\approx
  z^{\nu_{\pm}} \sqrt{Q_0/Q(z)}\; \times\nonumber\\
  &\exp\left(\pm\frac{1}{s}\int_0^z
    \frac{\text{d}t}{t} 
    \left[\sqrt{ Q^2(t)} - Q_0\right] \right). \label{LangerCorrectedWKBApproximation}
\end{align}
Here $Q^2(z) = \frac{1}{4}s^2 (\nu_+-\nu_-)^2 + \sum_{n=0}^N \text{v}_n z^{n+1}$,
and $Q_0 = Q(0)$. 
In equation (\ref{LangerCorrectedWKBApproximation}) we distinguish
between the $\nu_+$- and $\nu_-$-solutions, because the difference $\nu_+ - \nu_-$ may
in principle be large. 

The WKB integrals must in general be done numerically, sometimes along curves
in the complex plane. This requires careful attention to branch cuts. Here
we will only give some examples where most of the calculations can be done
analytically.

\subsection{Example 1: Anharmonic oscillators}

Consider the equation
\begin{equation}
   -\frac{\partial^2}{\partial y^2}\Psi(y) + \left(y^2+ c^2\right)^2\Psi(y) = 0,
   \label{Anharmonic_oscillator}
\end{equation}
for  real $c$ so that $c^2 \ge 0$. For large $y$ the typical solution behaves like
\begin{equation}
     \Psi(y) \sim \text{e}^{\frac{1}{3}y^3 + c^2 y},
     \label{CrudeWKBApproximation}
\end{equation}
neglecting the slowly varying prefactor. For a given value of $\vert y \vert$
this is maximum along the positive real axis.
Hence, with $x=y^2=\text{e}^{u}$, we find as a leading approximation
\begin{equation*}
     S(u) = {\textstyle \frac{1}{3}}\left(\text{e}^{\frac{3}{2}u} + 3 c^2 \text{e}^{\frac{1}{2}u}\right).
\end{equation*}
In this case the Frobenius series can be written
\begin{equation}
  \Psi(y) = \sum_{m=0}^{\infty} a_m\, y^{2m + \nu} \equiv \sum_{m=0}^{\infty} A_m(y),
\end{equation}
with $\nu=0,\;1$.
Ignoring the $\log(S''(u))$-term in~(\ref{mExpression}, \ref{sExpression}, \ref{logCorrectedS0})
we find
\begin{align}
  {m} &= {\textstyle \frac{1}{2}}\left(\text{e}^{\frac{3}{2}u} + c^2\,\text{e}^{\frac{1}{2}u} \right),
  \label{m_Anharmonic}
  \\
  \log\left(\left| a_{{m}}\right|\right) &= 
  \left({\textstyle \frac{1}{3}} - {\textstyle \frac{1}{2}} u\right) \text{e}^{\frac{3}{2}u} + 
  c^2 \left(1 -{\textstyle \frac{1}{2}} u \right) \text{e}^{\frac{1}{2}u}.
  \label{a_m_Anharmonic}
\end{align}

\begin{figure}[!t]
\begin{center}
\includegraphics[clip, trim = 8ex 6ex 9ex 5ex, width=0.483\textwidth]{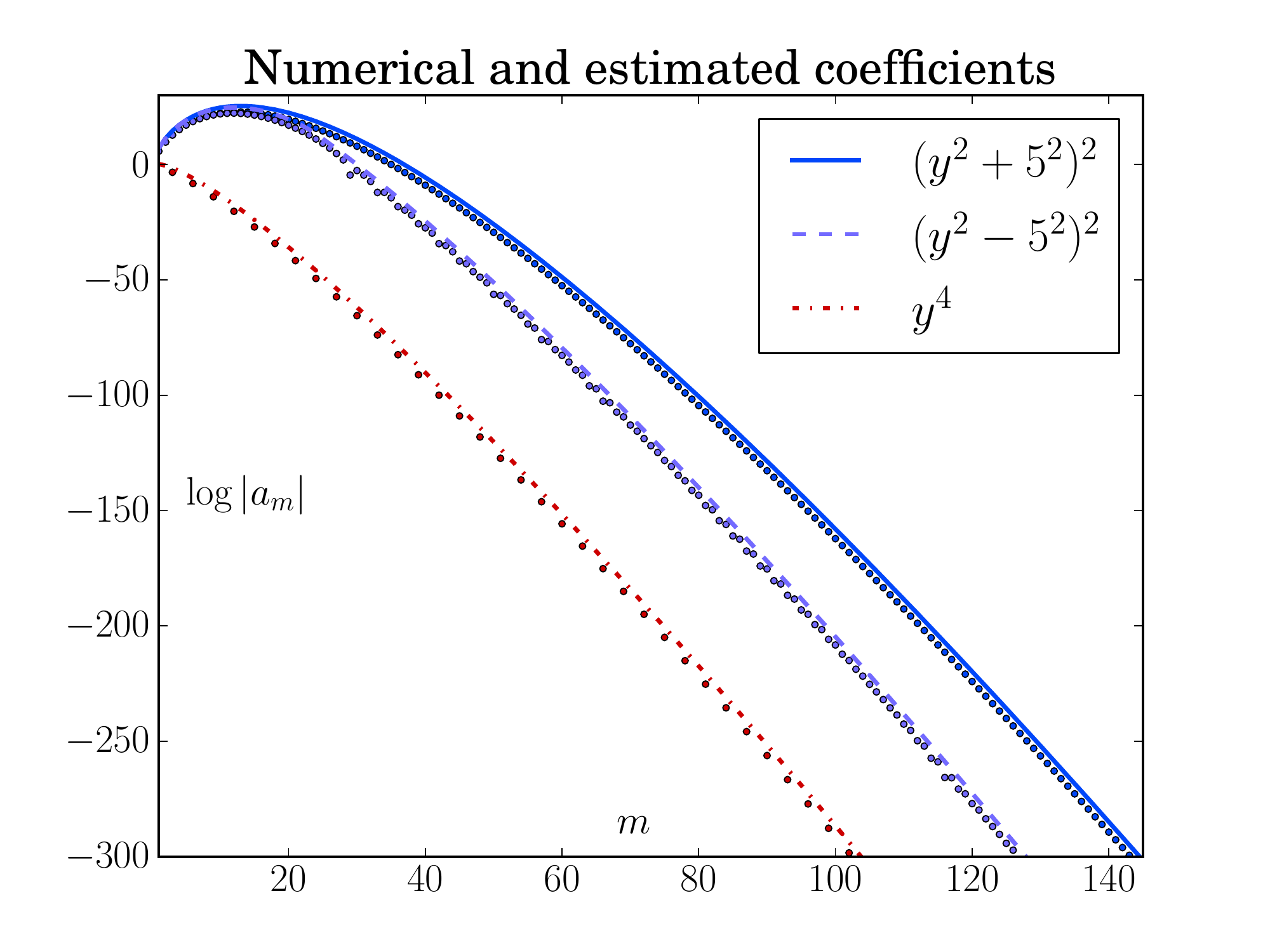}
\end{center}
\caption{Comparison of numerical coefficients $a_m$ (points)
with estimates (full-drawn lines) based on
(\ref{m_Anharmonic}, \ref{a_m_Anharmonic}) and
(\ref{m_DoubleWell}, \ref{a_m_DoubleWell}). The estimates
of $\log\vert a_m \vert$ are accurate up to corrections which
depend logarithmically on $m$.
}
\label{Coefficients_a_m}
\end{figure}

\noindent
For $c=0$ an explicit representation is
\begin{equation}
   \log \vert a_m \vert = \frac{2}{3}m\left(1 -\log 2m \right).
   \label{prediction0}
\end{equation}
This is plotted as the lower curve in figure~\ref{Coefficients_a_m}. It fits satisfactory
with the high-precision coefficients generated numerically, but there remains a
correction which depends logarithmically on $m$. For nonzero $c$ the parametric
representation provides equally good results, as shown by the upper curve in 
figure~\ref{Coefficients_a_m}.

\begin{figure}[!t]
\begin{center}
\includegraphics[clip, trim = 10.5ex 5ex 10ex 5ex, width=0.483\textwidth]{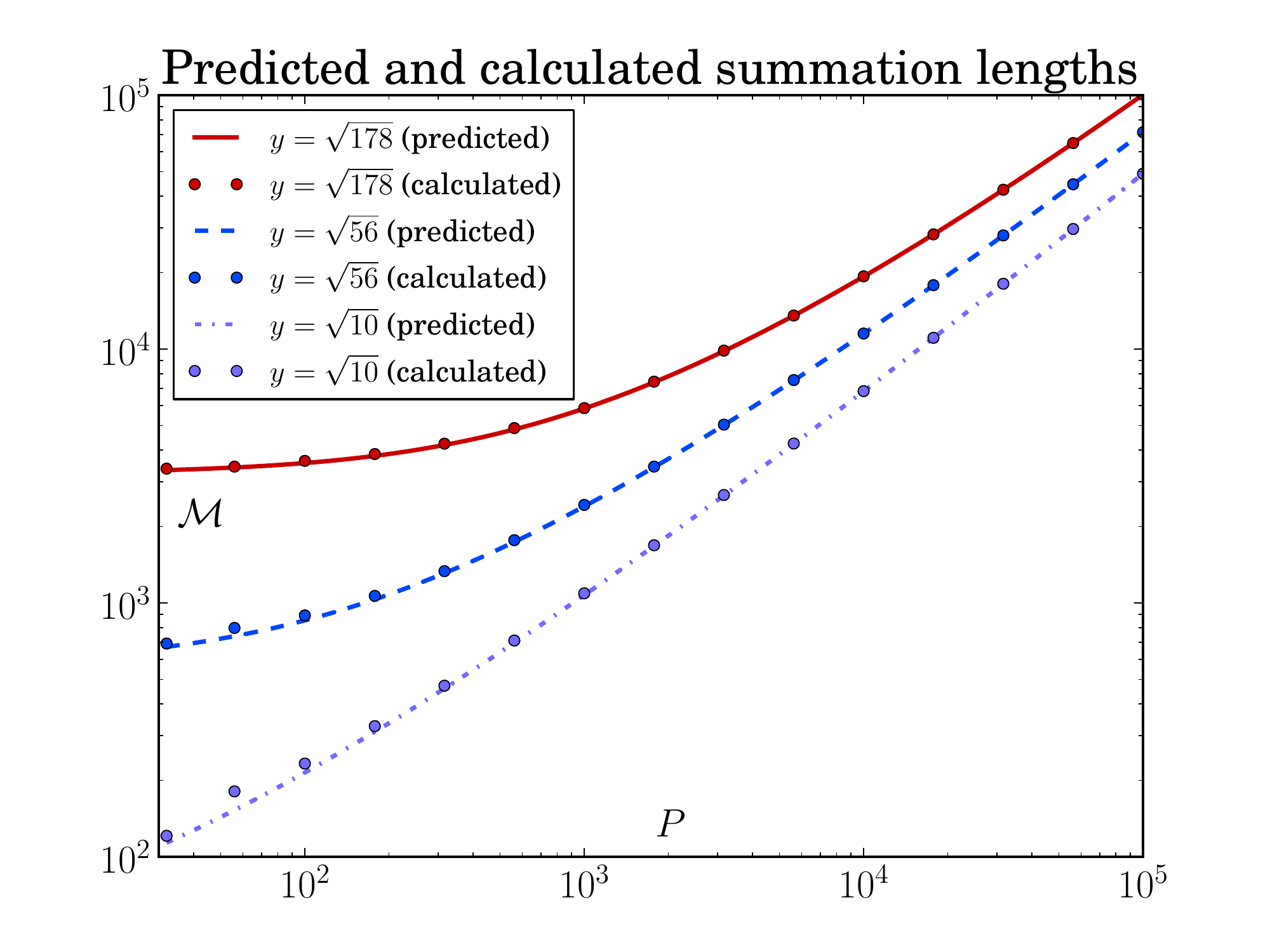}
\end{center}
\caption{This figure compares the {\em a priori\/} prediction,
based on equation~(\ref{prediction0}),
of the number of terms ${\cal M}$ which must be summed in order to evaluate
$\Psi(y)$ for $c=0$ to a desired precision $P$ with the actual number of terms
computed by.
}
\label{lengthOfSums}
\end{figure}

The conclusion of this example is that for a fixed (large) $x$
we expect the largest term of the power series to be
\begin{equation}
    \mathop{\text{max}}_m \vert A_m(x) \vert \sim \text{e}^{\frac{1}{3}(x^{3/2}+3 c^2 x^{1/2})},
\end{equation}
neglecting a slowly varying prefactor.
Further, the maximum should occur at
\begin{equation}
    m \approx {\textstyle \frac{1}{2}} \left( x^{3/2} + c^2 x^{1/2} \right).
\end{equation}
Finally, estimates like equation (\ref{prediction0})
for the coefficients $a_m$ may be used to predict how many terms ${\cal M}$
we must sum to evaluate $\psi(x)$ to a given precision $P$,
based on the stopping criterium
\begin{equation}
       \vert a_{\cal M} \vert\, x^{\cal{M}} \le 10^{-P}.
\end{equation}
As can be seen in figure~\ref{lengthOfSums} the agreement with the actual
number of terms used by our evaluation routine is good,
in particular for high precision $P$. But keep in mind that a
logarithmic scale makes it easier for a comparison to look good.

\noindent
Next consider the logarithmic corrections. Including the prefactor of
equation~(\ref{CrudeWKBApproximation}) changes $S(u)$ by an amount
\begin{equation}
  \Delta S(u) = -\frac{1}{2}\log\left(\text{e}^u + c^2\right).
\end{equation}
Including the $\log(S''(u))$-term in the relation between
$S(u)$ and $S_0(u)$ changes $S_0$ by an additional amount
\begin{equation}
   \Delta S_0(u) = 
   -\frac{1}{2}\log\left(\frac{3}{4}\text{e}^{\frac{3}{2}u}+\frac{1}{4}c^2\,\text{e}^{\frac{1}{2}u}\right).
\end{equation}
For $c^2 = 0$ this changes the relation~(\ref{prediction0}) to
\begin{equation}
   \log \vert a_m \vert = \frac{1}{3}\left(2m+{5}/{2}\right)
   \left(1 -\log \left(2 m + {5}/{2}\right)\right).
   \label{prediction1}
\end{equation}
For $\vert a_m \vert$ this essentially corresponds to a factor $ m^{-5/6}$.

\begin{figure}[!t]
\begin{center}
\includegraphics[clip, trim = 10.5ex 5ex 10ex 5ex, width=0.483\textwidth]{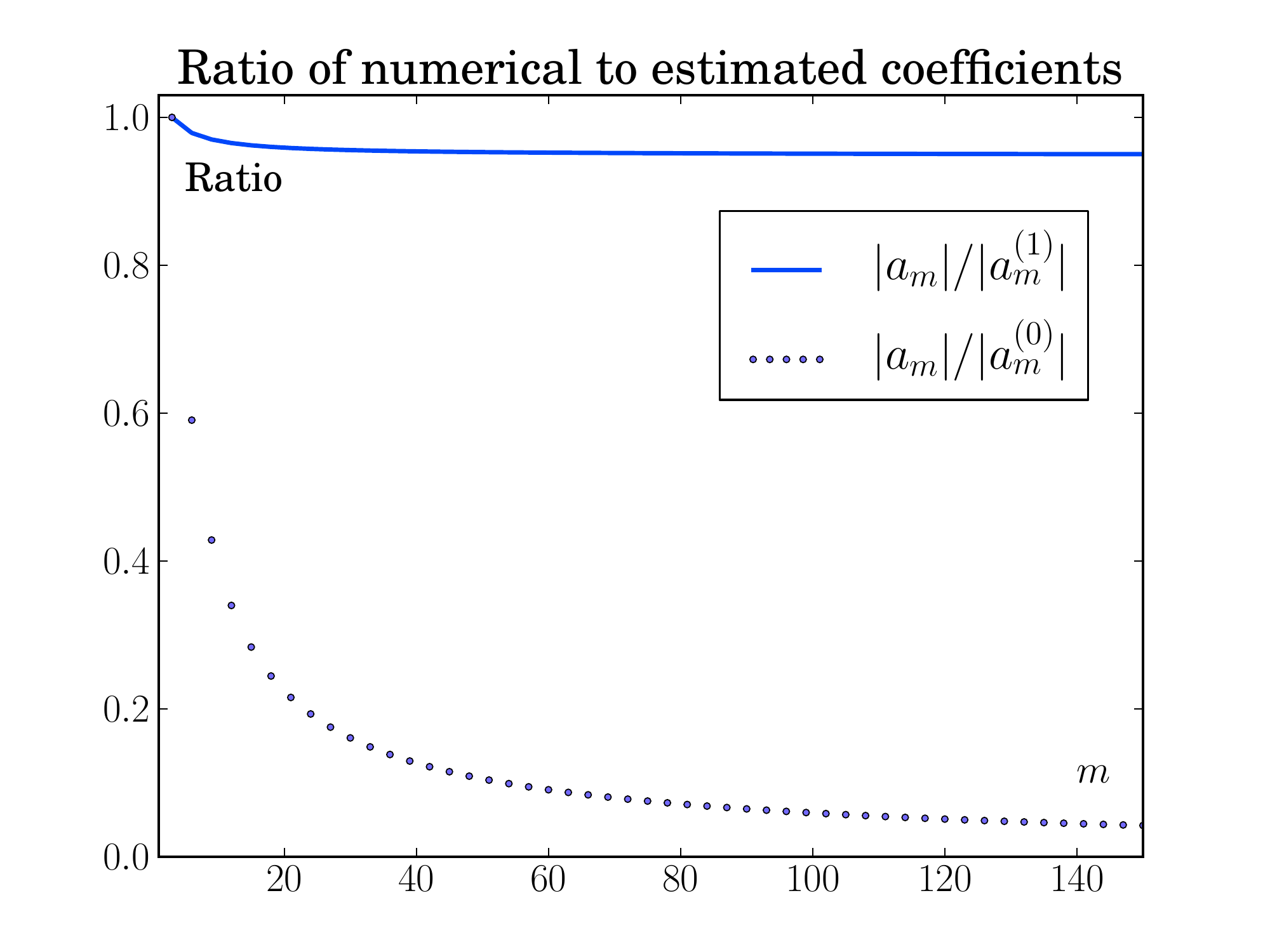}
\end{center}
\caption{This figure shows the ratio between the computed coefficients $a_m$ and the
crude prediction~(\ref{prediction0}) (labelled $\vert a^{(0)}_m \vert$) and the
logarithmically corrected prediction~(\ref{prediction1}) (labelled $\vert a^{(1)}_m \vert$).
For easy comparison we have in both cases adjusted an overall constant such that the ratio
is unity for $m=3$.   
}
\label{Ratios_a_m}
\end{figure}

\subsection{Example 2: Double well oscillators}

The same procedure also work for the equation
\begin{equation}
   -\frac{\partial^2}{\partial y^2}\Psi(y) + \left(y^2- c^2\right)^2\Psi(y) = 0,
   \label{Double_well}
\end{equation}
which however is a little more challenging since the maximum value of
$\vert\Psi(y\text{e}^{\text{i}\varphi}\vert$ sometimes occur for
$\varphi \ne 0$, i.e.~for complex arguments.

For large $y$ the typical solution behaves like
\begin{equation}
     \Psi(y) \sim \text{e}^{\frac{1}{3}y^3 - c^2 y},
\end{equation}
neglecting the slowly varying prefactor. Equation (\ref{Double_well})
can be transformed to the form (\ref{ODE}) by introducing $x=y^2$, 
$\Psi(y) = \psi(x)$. Hence, with $x=y^2=\text{e}^{u}$
\begin{equation*}
     S(u) = \mathop{\text{max}}_\varphi  {\textstyle \frac{1}{3}}\text{Re} \left(\text{e}^{\frac{3}{2}(u+\text{i}\varphi)} 
       - 3 c^2 \text{e}^{\frac{1}{2}(u+\text{i})\varphi}\right).
\end{equation*}
The maximum occurs for $\cos\frac{1}{2}\varphi = -\frac{1}{2}\left(1 + c^2\,\text{e}^{-u} \right)^{1/2}$ when
$\text{e}^{u} \ge \frac{1}{3} c^2$, and for $\cos\frac{1}{2}\varphi=-1$ otherwise.
This gives
\begin{equation}
     S(u) = \left\{\begin{array}{cc}
       {\textstyle c^2 \text{e}^{u/2} - \frac{1}{3}\text{e}^{3u/2}}&\text{for $e^u \le \frac{1}{3}c^2$,}\\[0.5ex]
       {\textstyle \frac{1}{3}} (\text{e}^{u} + c^2)^{3/2}&\text{for $e^u \ge \frac{1}{3}c^2$.}
       \end{array}
       \right.
\end{equation}
This implies that
{\footnotesize
\begin{align}
     \bar{m} &= 
     \left\{\begin{array}{lc}
         \frac{1}{2} \text{e}^{u/2}\left(c^2 - e^u\right)
         &\text{for $e^u \le \frac{1}{3}c^2$,}\\[0.5ex]
     {\textstyle \frac{1}{2}} \text{e}^{u}\,\left( \text{e}^u + c^2 \right)^{1/2}&\text{for $e^u \ge \frac{1}{3}c^2$},
     \end{array}
     \right.
     \label{m_DoubleWell}
     \\
     \log\left(\left| a_{\bar{m}}\right|\right) &= 
     \left\{\begin{array}{cc}
         \left(1\!-\!\frac{1}{2}u\right)c^2 \text{e}^{u/2} -\left(\frac{1}{3}-\frac{1}{2}u\right)\text{e}^{3u/2}
        &\text{for $e^u \le \frac{1}{3}c^2$,}\\[0.5ex]
     \left[\left({\textstyle \frac{1}{3}}\! -\!{\textstyle \frac{1}{2}}u \right)\text{e}^{u} 
       +{\textstyle \frac{1}{3}}c^2\right]\left(\text{e}^u + c^2\right)^{1/2}&\text{for $e^u \ge \frac{1}{3}c^2$}.
     \end{array}
     \right.
     \label{a_m_DoubleWell}
\end{align}
}
This representation compares fairly well with the numerically generated coefficients,
as shown by the middle curve in figure~\ref{Coefficients_a_m}. However, in this case
the coefficients $a_m$ have a local oscillating behaviour. The representation
(\ref{m_DoubleWell}, \ref{a_m_DoubleWell}) should be interpreted as the local amplitude
of this oscillation.

The conclusion of this example is that we expect the largest term of the power series to be
term of the series to be
\begin{equation}
    \mathop{\text{max}}_m \vert A_m(x) \vert \sim \text{e}^{\frac{1}{3}(x + c^2)^{3/2}},
\end{equation}
neglecting the slowly varying prefactor.
Further, the maximum should occur at
\begin{equation}
    m \approx {\textstyle \frac{1}{2}} x \left( x + c^2 \right)^{1/2} 
    \approx {\textstyle \frac{1}{2}} x^{3/2} + {\textstyle \frac{1}{4}} c^2 x^{1/2}.
\end{equation}

\section{Conclusion}

As illustrated in this contribution the coefficients of Frobenius series
can be predicted to surprisingly high accuracy by use of Legendre transformations
and lowest order WKB approximations. We have also tested the validity of the
method on many other cases.

%


%

\appendices

\section*{Acknowledgment}

We thank A.~Mushtaq and I.~{\O}verb{\o} for useful discussions.
This work was supported in part by the Higher Education
Commission of Pakistran (HEC).

%

\ifCLASSOPTIONcaptionsoff
  \newpage
\fi



%

\end{document}